\documentclass[10pt,letterpaper]{article}
\usepackage{opex3}
\usepackage{color}
\usepackage{cite}
\usepackage{amssymb}
\usepackage{amsmath}
\usepackage{graphicx}
\begin{document}

\title{Influence of the substrate material on the knife-edge based profiling of tightly focused light beams}

\author{C. Huber,$^{1,2,*}$  S. Orlov,$^{1,2,3}$  P. Banzer,$^{1,2}$ and G. Leuchs$^{1,2}$ }

\address{$^1$Max Planck Institute for the Science of Light, G\"unther-Scharowsky-Str.1, D-91058, Erlangen, Germany\\
$^2$Institute of Optics, Information and Photonics, University Erlangen-Nuremberg, Staudtstr. 7/B2, D-91058 Erlangen, Germany\\
 $^3$Center for Physical Sciences and Technology, Savanoriu Ave. 231, LT-02300 Vilnius, Lithuania}
\email{$^*$christian.huber@mpl.mpg.de} %% email address is required

\begin{abstract}
The performance of the knife-edge method as a beam profiling technique for tightly focused light beams depends on several parameters, such as the material and height of the knife-pad as well as the polarization and wavelength of the focused light beam under study. Here we demonstrate that the choice of the substrate the knife-pads are fabricated on has a crucial influence on the reconstructed beam projections as well. We employ an analytical model for the interaction of the knife-pad with the beam and report good agreement between our numerical and experimental results. Moreover, we simplify the analytical model and demonstrate, in which way the underlying physical effects lead to the apparent polarization dependent beam shifts and changes of the beamwidth for different substrate materials and heights of the knife-pad. 
\end{abstract}

\ocis{(140.3295) Laser beam characterization; (260.5430) Polarization; (050.6624)
Subwavelength structures; (050.1940) Diffraction; (240.6680) Surface plasmons.} % REPLACE WITH CORRECT OCIS CODES FOR YOUR ARTICLE, MINIMUM OF TWO; Avoid using the OCIS codes for “General” or “General science” whenever possible.
%For a complete list of OCIS codes, visit: http://www.opticsinfobase.org/submit/ocis/

%%%%%%%%%%%%%%%%%%%%%%% References %%%%%%%%%%%%%%%%%%%%%%%%%

%%%%%%%%%%%%%%%%%%%%%%%%%%  body  %%%%%%%%%%%%%%%%%%%%%%%%%%
\section{Introduction}

Due to their complex and yet controllable field distributions, tightly focused laser beams are known as versatile tools for nano-optics, plasmonics and microscopy  \cite{PBan10,XWChe10,NHus01,PBan10a}. For such studies, the precise knowledge of the 'tool' itself is of particular importance. For such tightly focused and highly confined light beams, several beam reconstruction techniques have been proposed and discussed in literature. Some of these methods even allow for the measurement of amplitudes and phases of individual electric field components in diffraction-limited focal spots \cite{TBau14,TGro10,MBur09,MSch10}. Another technique, which is normally used to determine the electric field intensity distribution in the cross-section of light beams experimentally, is the so-called knife-edge method which we want to discuss here in more detail \cite{JAArn71,AHFir77,RDorn03a,RDorn03b,PMar11,CHub13}.

The operation principle of this method, adopted from the profiling technique of beams with diameters orders of magnitudes larger than the wavelength, is based on an opaque knife-pad or razor-blade which is line-scanned through the transverse cross-section of a beam under study. While scanning, the power of the transmitted light beam that is not blocked by the knife-pad is recorded by a detector. The beam profile can be tomographically reconstructed from the photo-current curves resulting from scans performed under different directions.

In an earlier communication we reported that in general the knife-edge method for tightly focused light beams may suffer from the interaction of the light beam with the knife-pad itself \cite{PMar11}. The shape of the reconstructed beam projections is distorted and their positions are shifted. The magnitude of such distortions strongly depends on several parameters, such as material parameters and height of the knife-pad as well as the wavelength and polarization of the beam under study. As a consequence, a proper and careful choice of these system parameters, especially of the knife-pad material, is crucial to allow for an accurate reconstruction \cite{RDorn03a}.

\begin{figure}[t!]
\centering\includegraphics[scale=0.55]{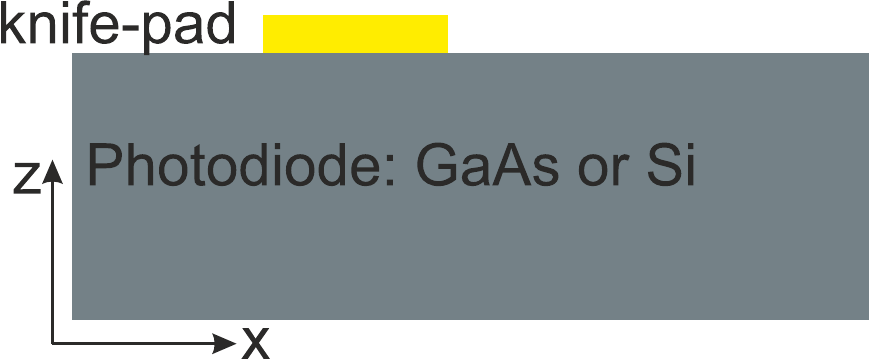} \text{(a)} 
\centering\includegraphics[scale=0.55]{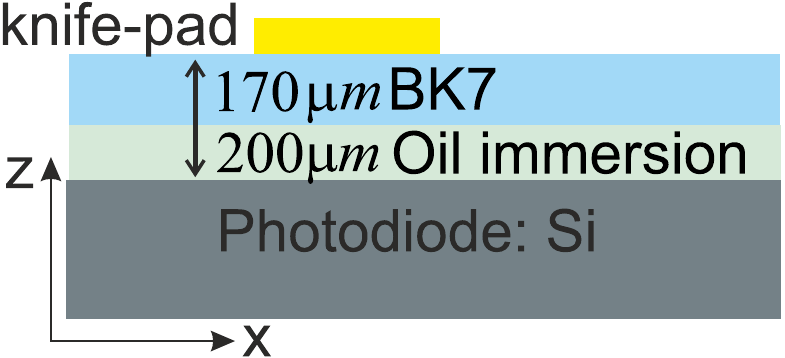} \text{(b)} 
\centering\includegraphics[scale=0.52]{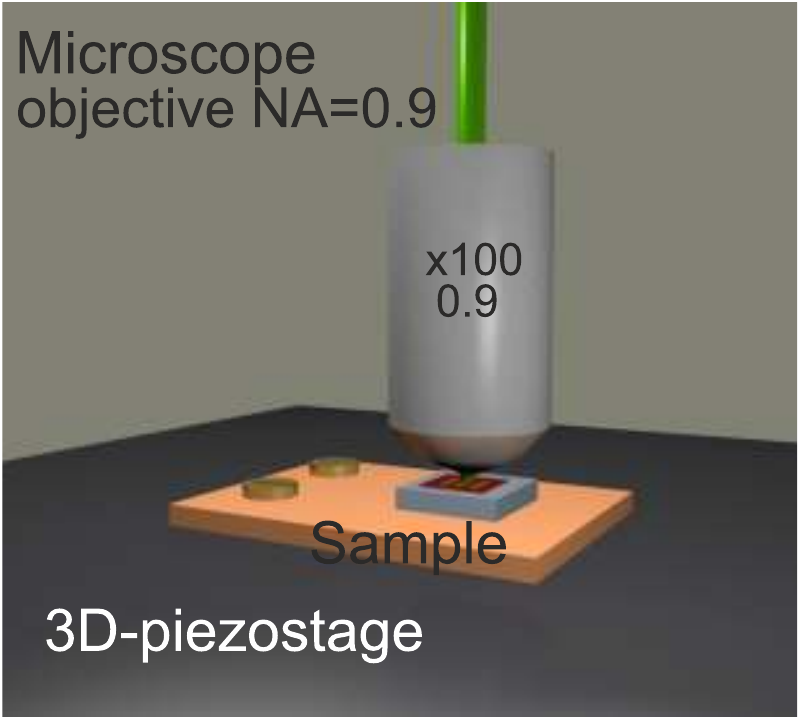} \text{(c)} 
\centering\includegraphics[scale=0.52]{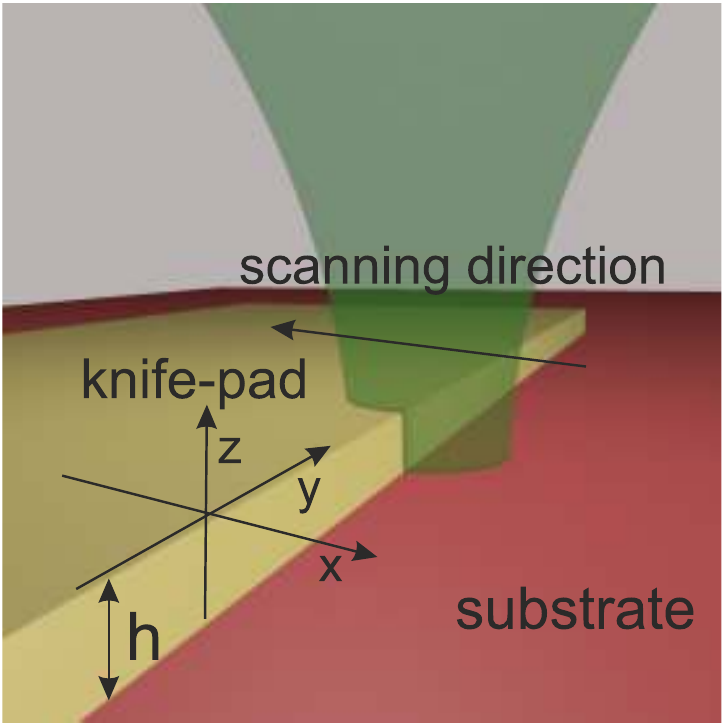} \text{(d)} 
\centering\includegraphics[scale=0.35]{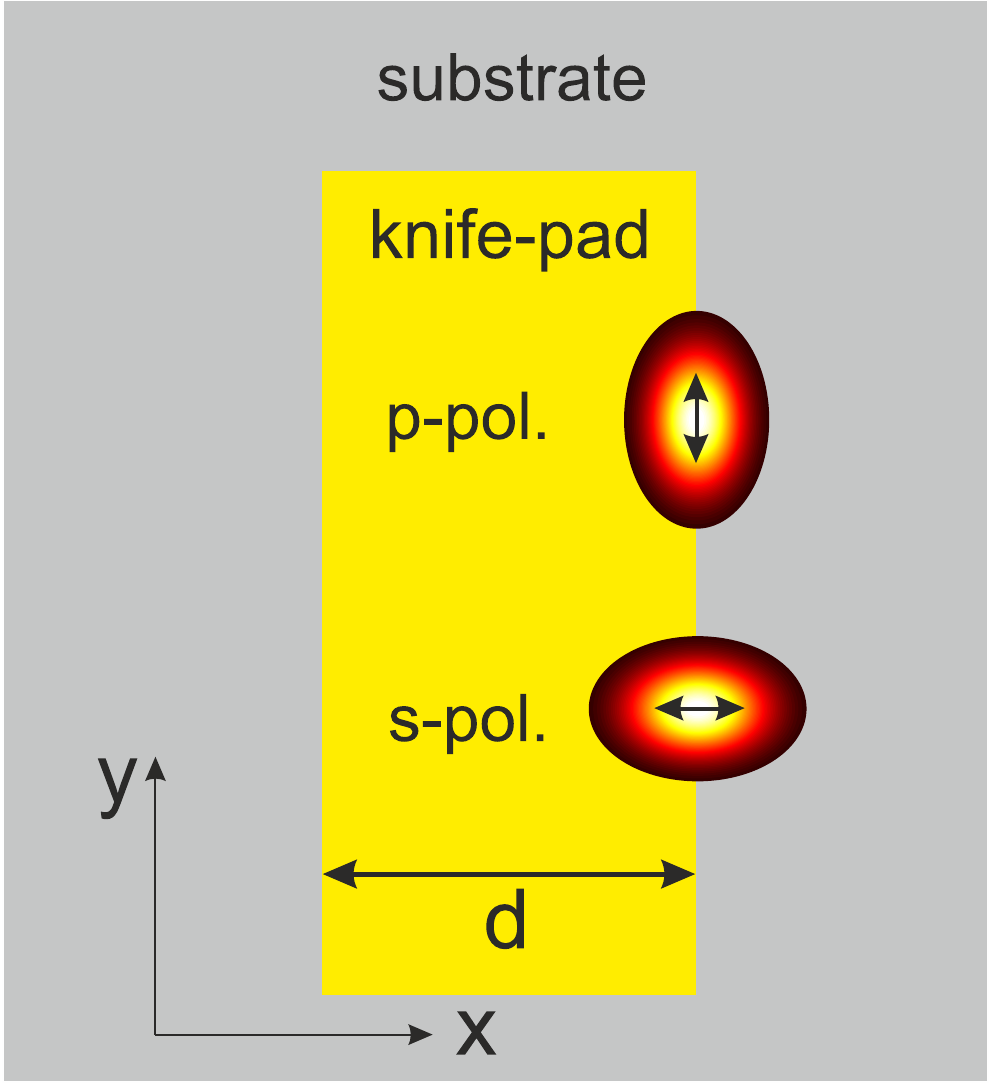} \text{(e)} 
\caption{Schematic sketch of the samples and the setup. The knife-pads are directly fabricated on GaAs- or Si-photodiodes (a) or manufactured on glass substrates (BK7) that are put on a Si-photodiode (size of the area: $6$ $x$ $6$ $\mu$m) with a thin layer of immersion oil in between (b). A linearly polarized Gaussian beam is focused on the samples by a high numerical aperture objective (NA 0.9). The samples are mounted on a holder that can be moved by a 3D-piezostage with nanometer accuracy (c). For the knife-edge measurements, the sample is moved through the focal spot and the transmitted light is detected by a photodiode underneath (d,e).}
\label{fig:setup}
\end{figure}

As mentioned already in earlier studies \cite{PMar11}, one dominant effect leading to distorted reconstruction data is the plasmonic excitation of metallic knife-pads, which depends on the polarization and wavelength of the input beam as well as on material properties and the dimensions of the knife-pad. In this context, it is known that the resonance behavior of, for instance, plasmonic particles is influenced drastically in case of a dielectric interface being placed in close proximity (see for instance \cite{MWKni09,Aug10,CJep10}). We now study the influence of the substrate material on the knife-edge profiling scheme in detail by performing knife-edge measurements for different detector and substrate materials.

Following up on our earlier works, we now also include these previously unaccounted effects into our theoretical treatment of the problem that are introduced by the material of the substrate, onto which  the knife-pads are fabricated. Our numerical calculations are mainly based on a modified version of the analytical model presented in Ref. \cite{PMar11}. Here, we have now removed the second knife-pad and include nonlocalised solutions \cite{BStu07}. Along with the exact analytical model we are introducing, for illustratory purposes, a largely simplified theoretical approach based on Ref.  \cite{CHub13}, where an alternative method for an accurate retrieval of the beam parameters was discussed, even for cases in which the conventional approach fails. We restrict ourselves in this simplified model to the assumption that the knife-pad does not only block the beam but also interacts with the local electric field and its gradients. The plausability of this assumption was largely confirmed and successfully employed to account for spurious effects introduced by the interaction of the beam with the knife-pad \cite{CHub13}.

\section{Setup and Samples}

For the measurement knife-pads made of gold with a height $h$ of $130$ nm and $70$ nm have been fabricated on top of two different types of photodiodes as well as on a glass substrate (BK7). As used in our former experiments, custom-built GaAs p-i-n photodiodes have been utilized as detector respectively substrate material for the knife-edge samples. In addition we now have also fabricated samples on two more substrates: silicon photodiodes and glass-substrates (BK7). In the latter case the glass substrate was placed on a photodiode for detection. Silicon photodiodes normally exhibit a thin SiO$_2$ protection layer, which in our case has a height of about $120$ nm. This value was measured using ellipsometry and is in line with the specifications provided by the manufacturer \cite{LCOMP}. In case of the glass substrate, immersion oil was used to fill the unavoidable, thin air gap between the substrate and the detector (also a silicon photodiode) underneath. Schematic sketches of the samples are depicted in Fig. \ref{fig:setup} (a), (b). Considering the dimensions of the silicon photodiode and the thickness of the oil immersion layer, an effective numerical aperture (NA) of about $1.48$ can be estimated for the detection system on glass. 

For the knife-edge measurement, the structures are scanned stepwise through the focal spot of a highly focused beam and the power of the light not blocked by the knife-pad is measured. A theoretical description of the knife-edge measurement can be found in Section 4. A schematic sketch of the setup is shown in Fig. \ref{fig:setup} (c). Using a high NA objective (NA$=0.9$ in air) the light is focused onto the samples which can be moved with nanometer accuracy by a piezo-stage. Within one line scan, both edges of the knife-pad that have a distance $d$ to each other are moved through the focal spot. The differentiated photocurrent curve corresponds to the projection of the beam profile which can easily be post-processed to retrieve the electric field intensity distribution in the focal plane \cite{SWRow79}, if the interaction of the beam with the knife-pad can be neglected. 
For a collimated linearly polarized Gaussian (TEM$_{00}$) laser beam used in the study presented here, the focal spot is elongated in the direction of the polarization of the incoming light beam \cite{RDorn03a}. To analyze the performance of the chosen knife-edge samples, we measure along the major and minor axes of the focal spot. Hence, we choose two scanning directions for our measurements, one along the axis which is parallel to the polarization of the incoming beam (incoming polarization perpendicular (s) to the knife-pad) and another perpendicular to this axis (incoming polarization parallel (p) to the knife-pad), see Fig.  \ref{fig:setup} (e). 
For a detailed discussion about the principles of the measurement, our experimental setup and the procedure of measurement itself we would like to refer the reader to our previous publications \cite{PMar11,CHub13}.
It is worth noting here, that the input beam sizes of the collimated laser beam used in these experiments were chosen to be different from those presented in our earlier studies.
Furthermore we are not using periodic strip-like structures anymore but individual knife-edge structures only, as already discussed in \cite{PMar11}. In this way a possible interaction between nearby structures can be completely excluded. Furthermore the use of a Savitzky-Golay smoothing algorithm for filtering the photocurrent data is no longer necessary because of improved measuring conditions. Due to a higher signal-to-noise ratio achieved for the experimental data, the shape of the beam projections can be measured even more precisely now.

\section{Experimental and numerical results}

\begin{figure}[t!]
\centering\includegraphics[scale=0.25]{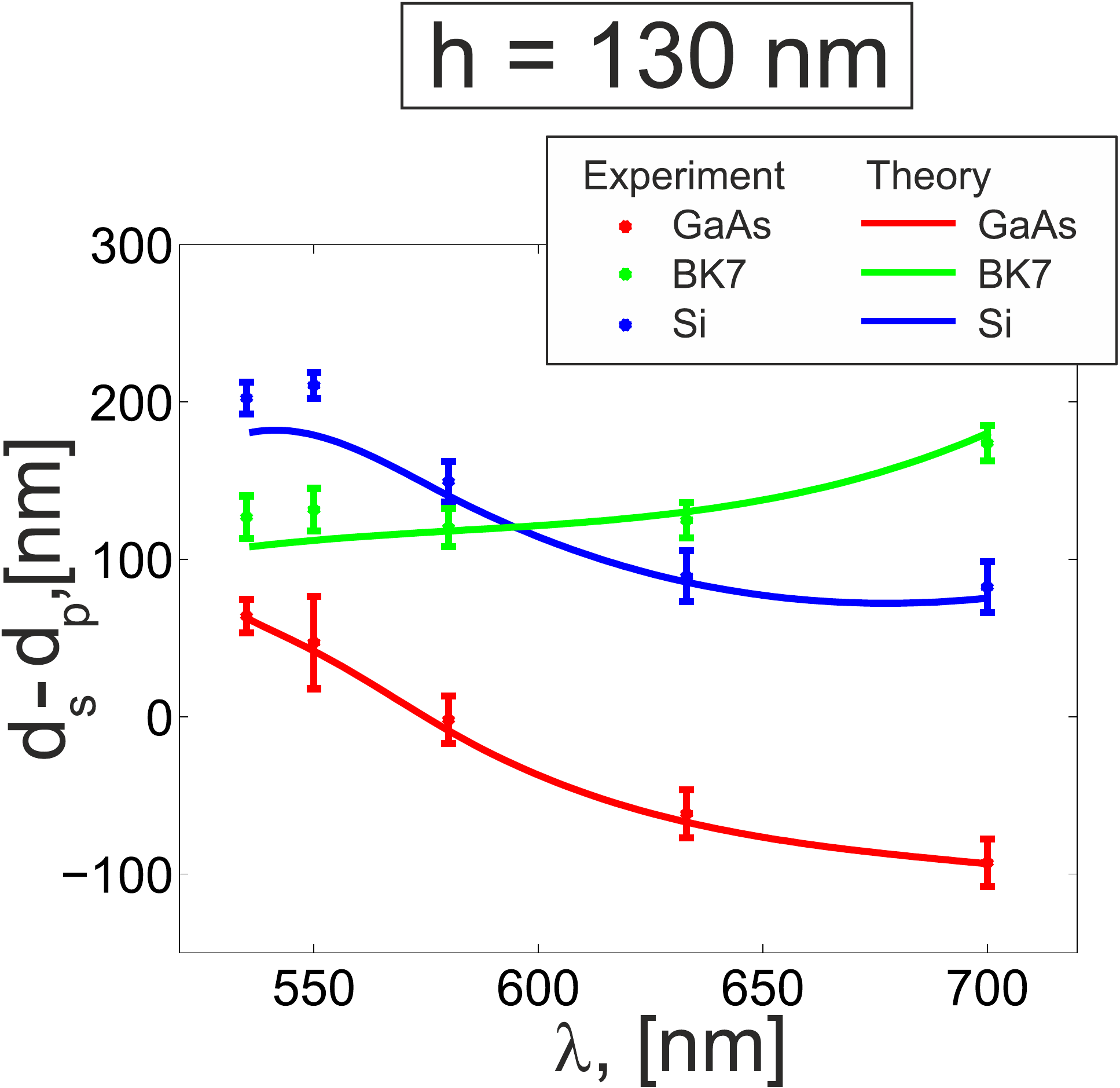} \text{(a)} 
\centering\includegraphics[scale=0.25]{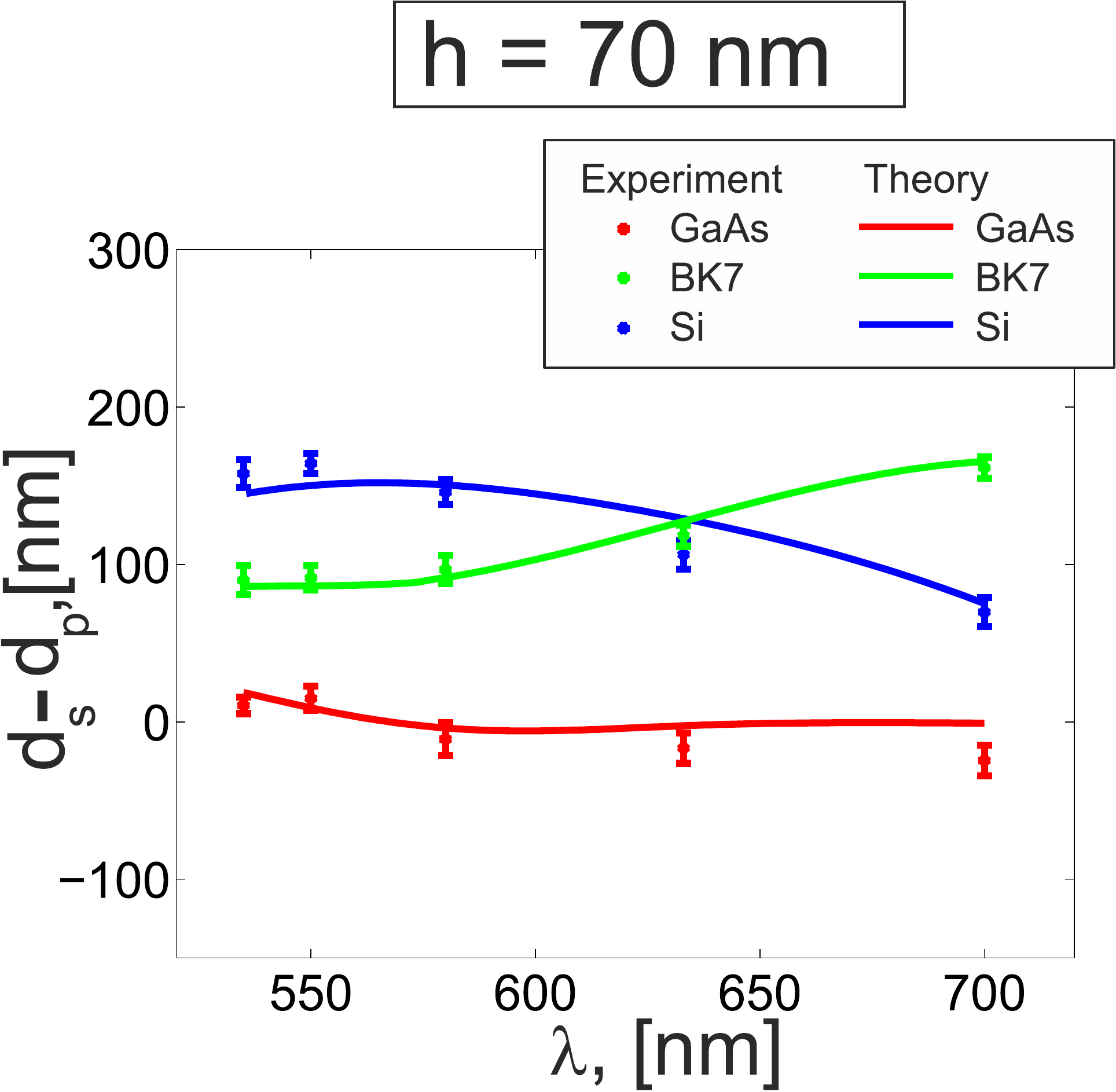} \text{(b)} 
\centering\includegraphics[scale=0.25]{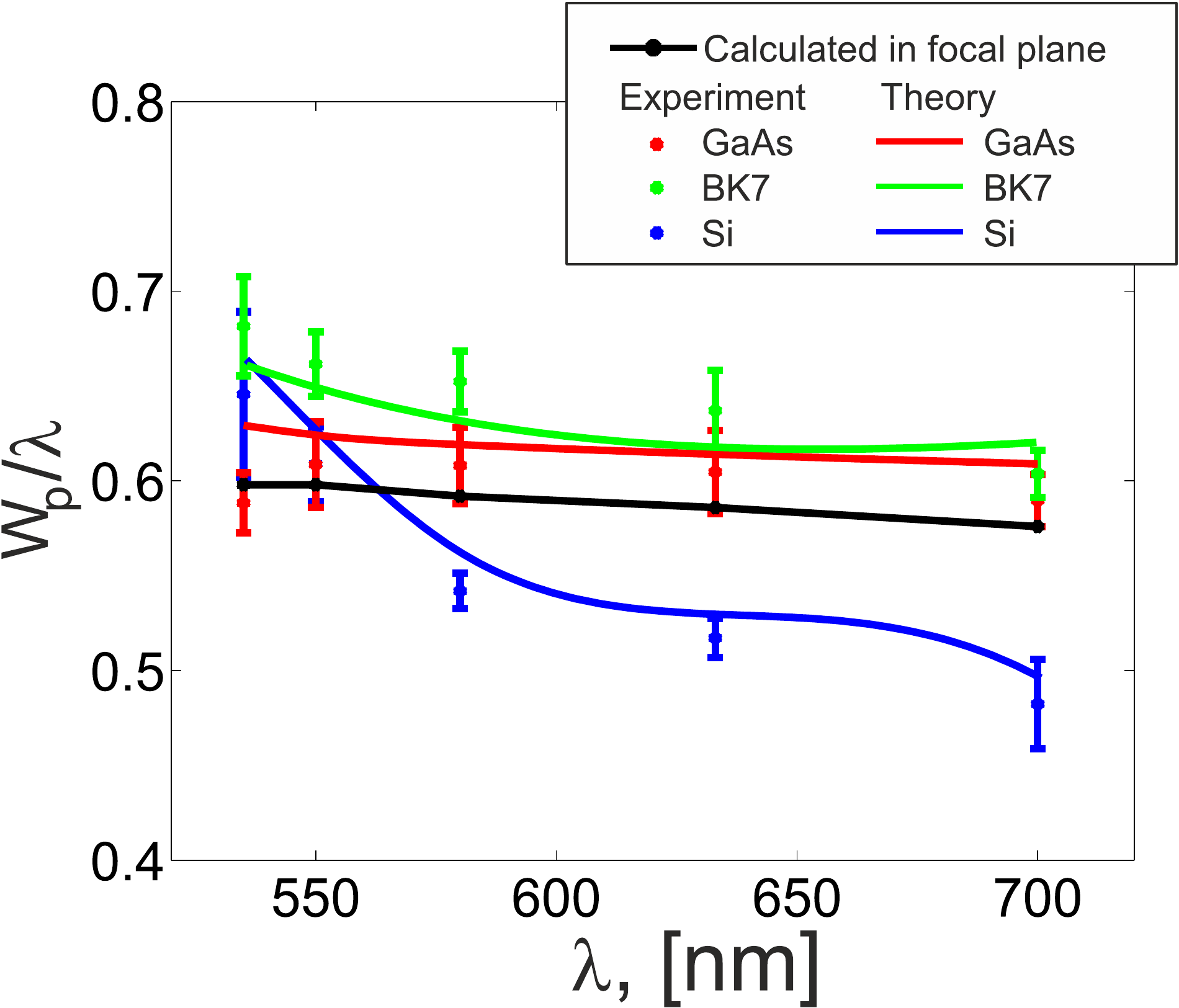} \text{(c)} 
\centering\includegraphics[scale=0.25]{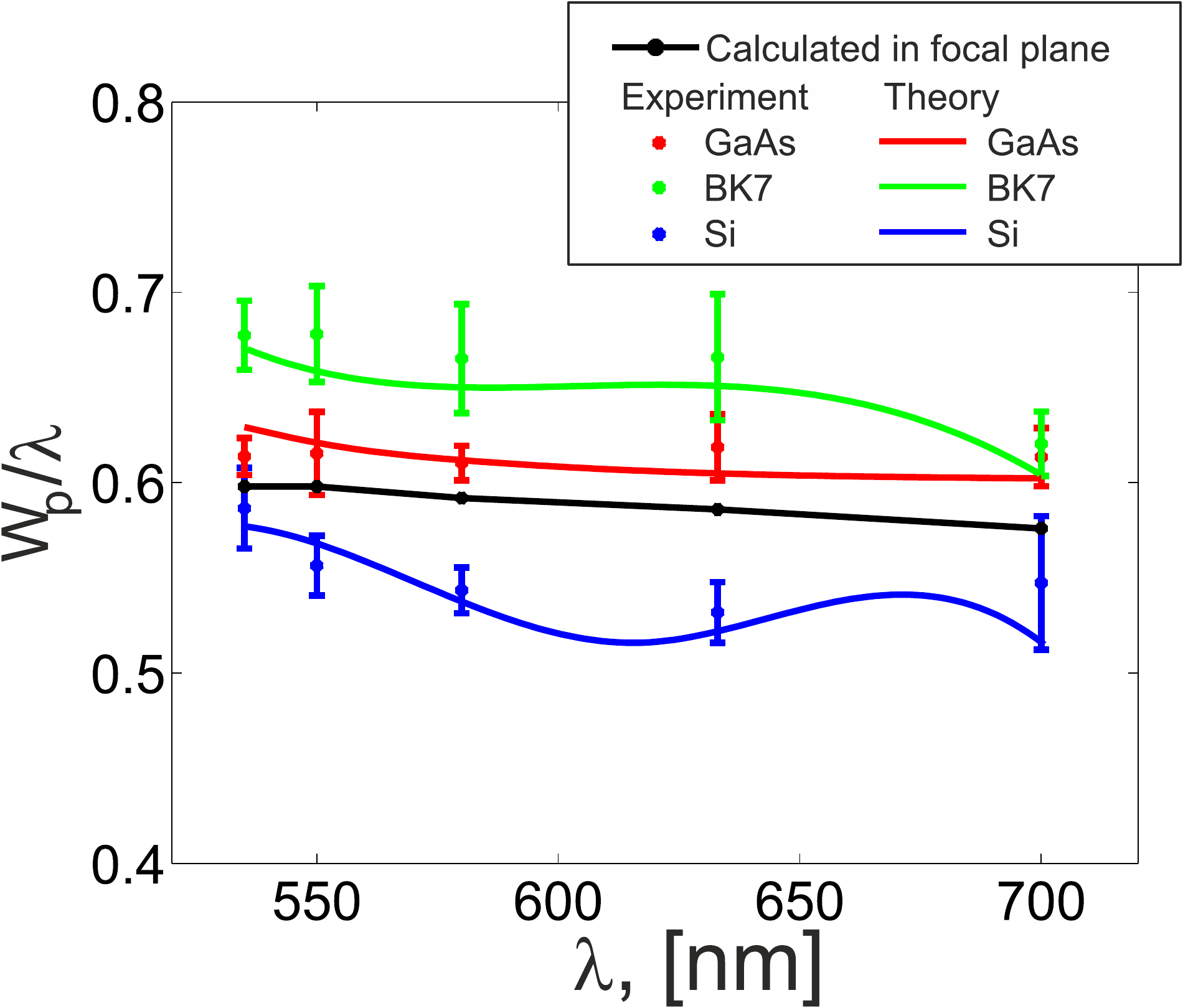} \text{(d)}
\centering\includegraphics[scale=0.25]{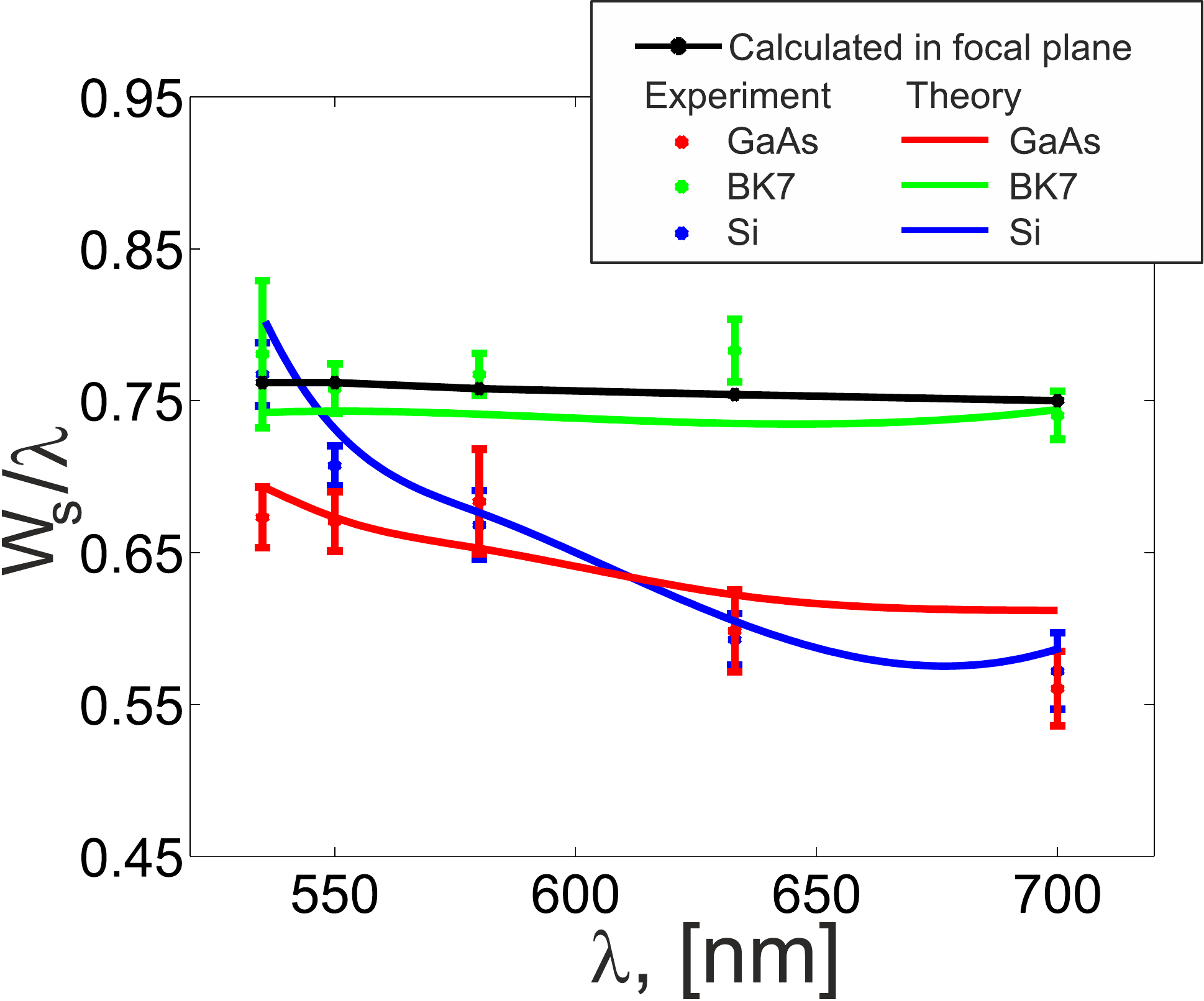} \text{(e)} 
\centering\includegraphics[scale=0.25]{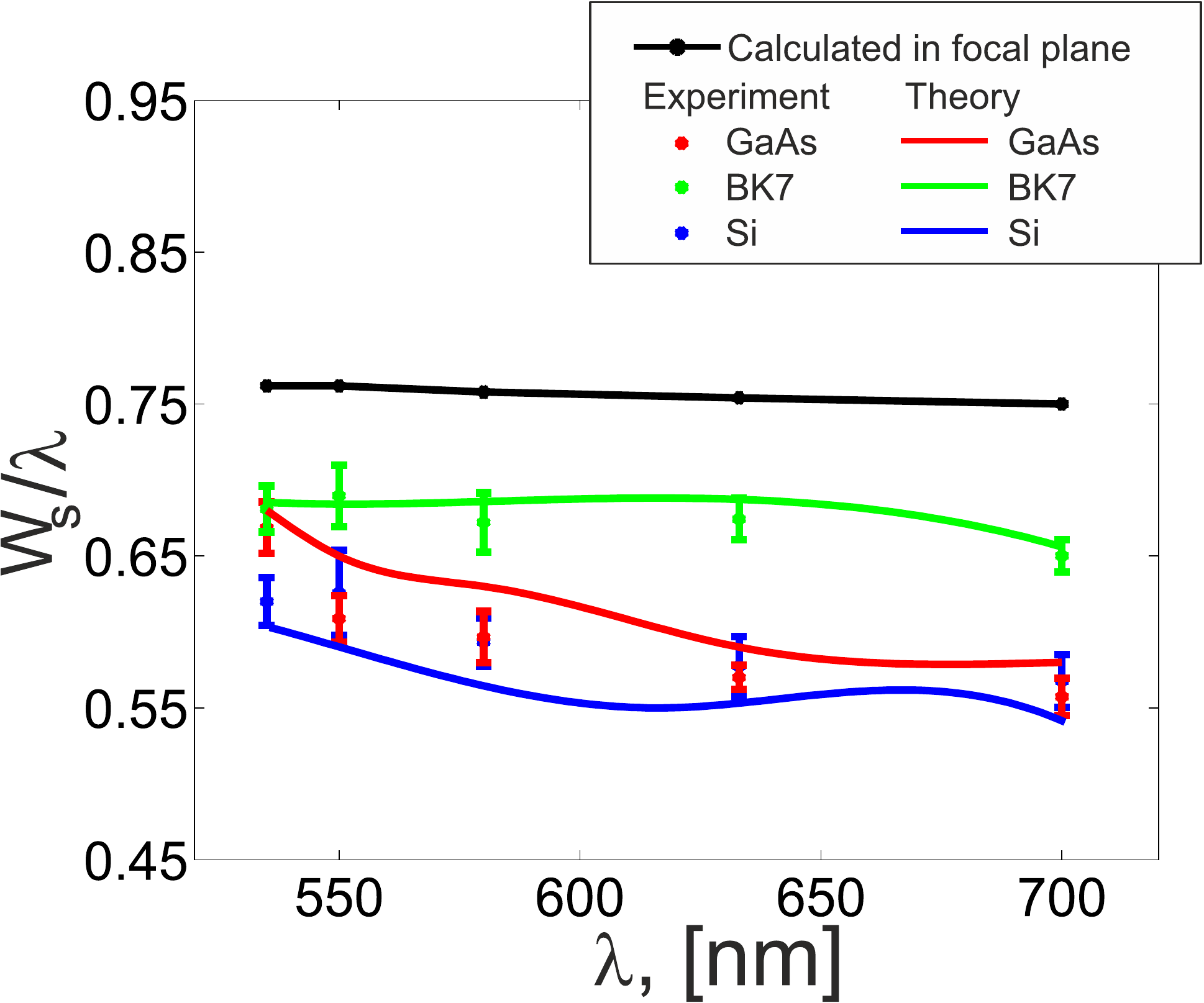} \text{(f)} 
\caption{Experimental and theoretical results plotted versus the wavelength for different substrate materials (GaAs-, Si-Diode and BK7 substrate) for Au knife-pads with a height $h=130$ nm (a,c,e) and $h=70$ nm (b,c,d). Shift $d_s-d_p$ between the maxima of the differentiated photocurrent curves (a, b). Reconstructed beam size of the focal spot for s- and p-polarized light $W_p$ (c,d) and $W_s$ (e,f) (each normalized to the wavelength). The calculated beams size in the focal plane is plotted in (b-f) as well.}
\label{fig:exp130_70}
\end{figure}

In this section we present the experimental results of our knife-edge measurements for two heights of the knife-pad and three different substrate materials as discussed above. The obtained projections which are modified by the interaction of the focal field with the knife-edge samples are characterized with respect to the parameters $d_s$, $d_p$, which correspond to the actual positions of the measured projections, and the FWHMs $W_s$  and  $W_p$ as described in Ref. \cite{PMar11}. Furthermore the asymmetric shape of the curves caused by a plasmonic excitation of the knife-pad will be shown and discussed within the article (see Chapter 4.3). As the shape of the measured projections slightly depend on the quality and roughness of the knife-pads, line-scans have been repeated at different positions along the knife-pads of each sample and average values and statistical errors are evaluated. The measured data are compared to numerical results obtained from our exact analytical model (see Section 4).
Both data-sets are in very good agreement with each other.

According to the theoretical model \cite{PMar11}, the measured beam profiles are shifted towards or away from the physical position of knife-pad, so that the values for $d_s$ and $d_p$ will not be equal to the width $d$ of the knife-pad (distance between both edges) in general. 
To avoid errors that come from an inaccuracy in determining the exact width of each knife-pad we analyze the relative shift $d_s-d_p$. These values for all substrate materials are shown for metal heights $h$ of  $130$ nm in Fig. \ref{fig:exp130_70} (a) and for  $70$ nm in Fig. \ref{fig:exp130_70} (b) respectively. 
Furthermore we find both experimentally and theoretically that for most of the cases the reconstructed beam sizes $W_s$ and $W_p$ are smaller or bigger than the actual beam size we have calculated in the focal plane. These values are presented in Fig. \ref{fig:exp130_70} (c-f) together with the results of the beam sizes calculated by vectorial diffraction theory \cite{PDeb09,EWol59,BRic59} (black line).  We note that these modifications are not caused by the focusing system, since we can independently characterize our focal spots using an alternative technique \cite{TBau14} but by the interaction of the beam and the knife-edge samples. 
In all figures the experimental points are compared to our theoretical calculations (continuous lines). 

From these results we can conclude that the measured values of the shift $d_s-d_p$ and the reconstructed beam sizes strongly differ from each other for different substrate materials for both pad heights. 
For a height of $130$ nm, we find as a characteristic feature that the sign of the shift $d_s-d_p$ is positive throughout the investigated spectral range for samples on Si and BK7 while for samples on GaAs this parameter also turns negative. Furthermore for samples on glass, the shift gets monotonically larger with increasing wavelength while this value is mostly decreasing for samples on GaAs and Si.
For a height of $70$ nm the situation is similar.

We also find that the measured values for $W_s$ are smaller than or close to the calculated beam widths for all substrates while $W_p$ is larger than the calculated beam width for GaAs and BK substrates. For samples on Si-diodes, $W_p$ is in general smaller than the calculated beam width but crosses the calculated beam size for shorter wavelengths regarding an a height of 130 nm (see Fig. \ref{fig:exp130_70} (c).

\section{Simplified theoretical model of the knife-edge method}

In our theoretical discussion we present a simplified model by which the occurring modifications of the beam projections can be explained. From that we can see how the substrate influences the underlying physical effects in particular. 
We start the discussion with a brief reminder of the knife-edge method basics. In the original work \cite{JAArn71}, the following assumptions were made: the incident beam is paraxial, the knife-pad is made from a perfect conductor and no losses are present. The photocurrent recorded by a detector is proportional to the power $P$ not blocked by a knife-pad and is recorded for each beam position $x_0$ with respect to the knife-pad 
\begin{equation}
 P=P_0\int _{-\infty}^{\infty}\mathrm{d}y \int _{0}^{\infty}  I\left(x+x_0,y,z=0 \right) \mathrm{d}x,
\label{eq:knife}
\end{equation}
where $P_0$ is a proportionality coefficient and $I$ is the total electric energy density distribution scanned by the knife-pad. In the conventional knife-edge method $P$ is proportional to the intensity of the beam not blocked by the knife-pad (i.e. the integration is performed from the knife-pad at $x=0$ to $\infty$). Therefore the derivative $\partial P/\partial x_0$ of the photocurrent curve with respect to the beam position $x_0$ reconstructs a projection of the intensity $I$ onto the $xz$-plane at $z=0$ (projection onto the x-axis) \cite{AHFir77}. 

Recently it was demonstrated that for tightly focussed beams at the nanoscale Eq. (\ref{eq:knife}) can be recast in terms of the projection of the total electric energy density distribution $U_E(x)$ onto the $xz$-plane at the knife-pad and its derivatives \cite{CHub13}:
\begin{align}
	\frac{\partial P}{P_0\partial x_0}=C_0U_{E}(x_0) + \sum_{n=1}^{\infty}C_n\frac{\partial^n U_{E}( x_0)}{\partial x_0^n}.
	\label{eq:adopt_der}
\end{align}
where the coefficients $C_n$ determine the specific knife-edge sample. In the conventional knife-edge method \cite{JAArn71} these coefficients could be neglected and have been set to be $C_n=0$, for $n>1$. We note that the change in the scan direction will result in new $C'_n$ coefficients in Eq. (\ref{eq:adopt_der}), which are related to the former ones by $C'_n=C_n (-1)^{n+1}$.

\begin{figure}[t!]
\centering
\includegraphics[scale=0.23]{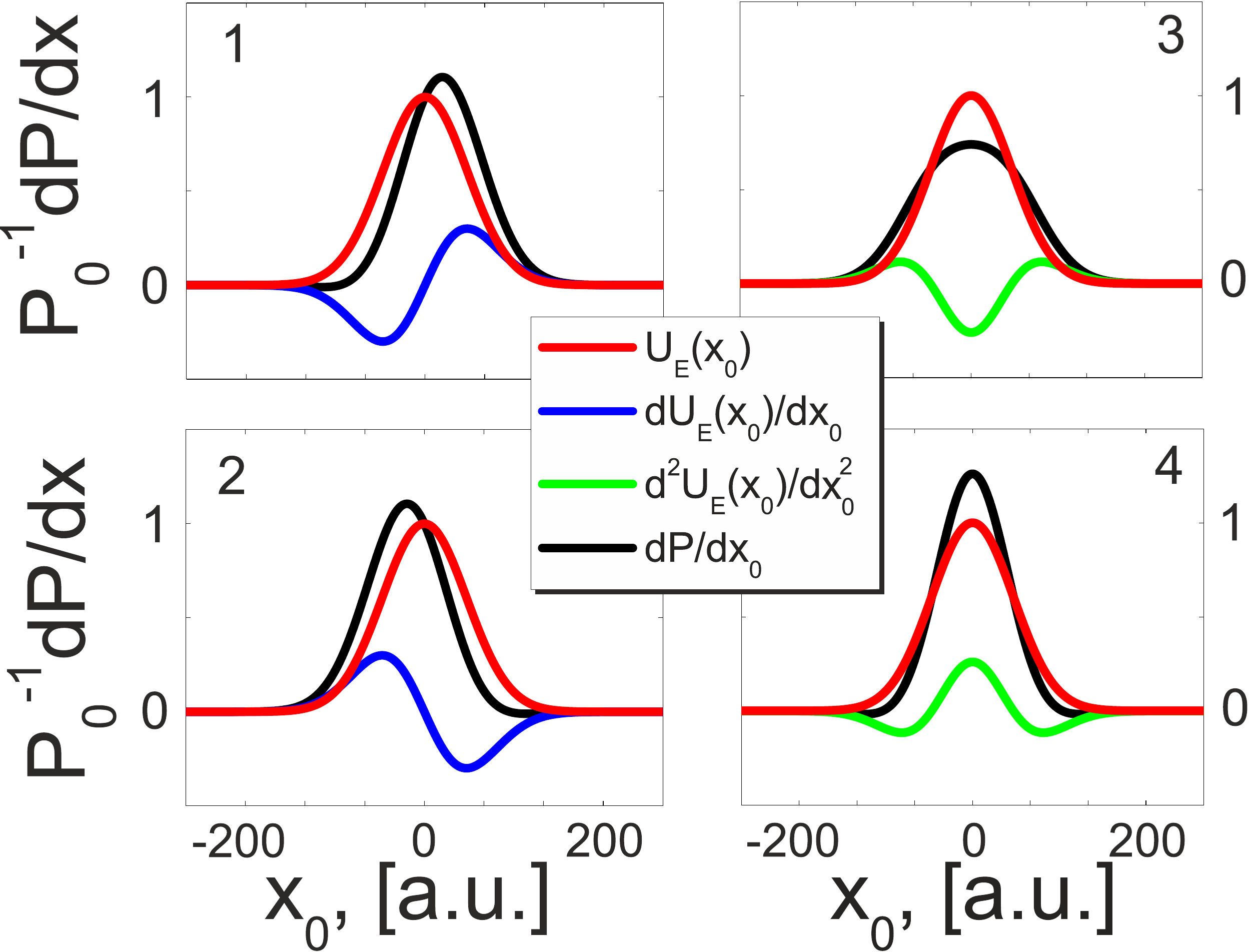} \text{(a)} 
\includegraphics[scale=0.23]{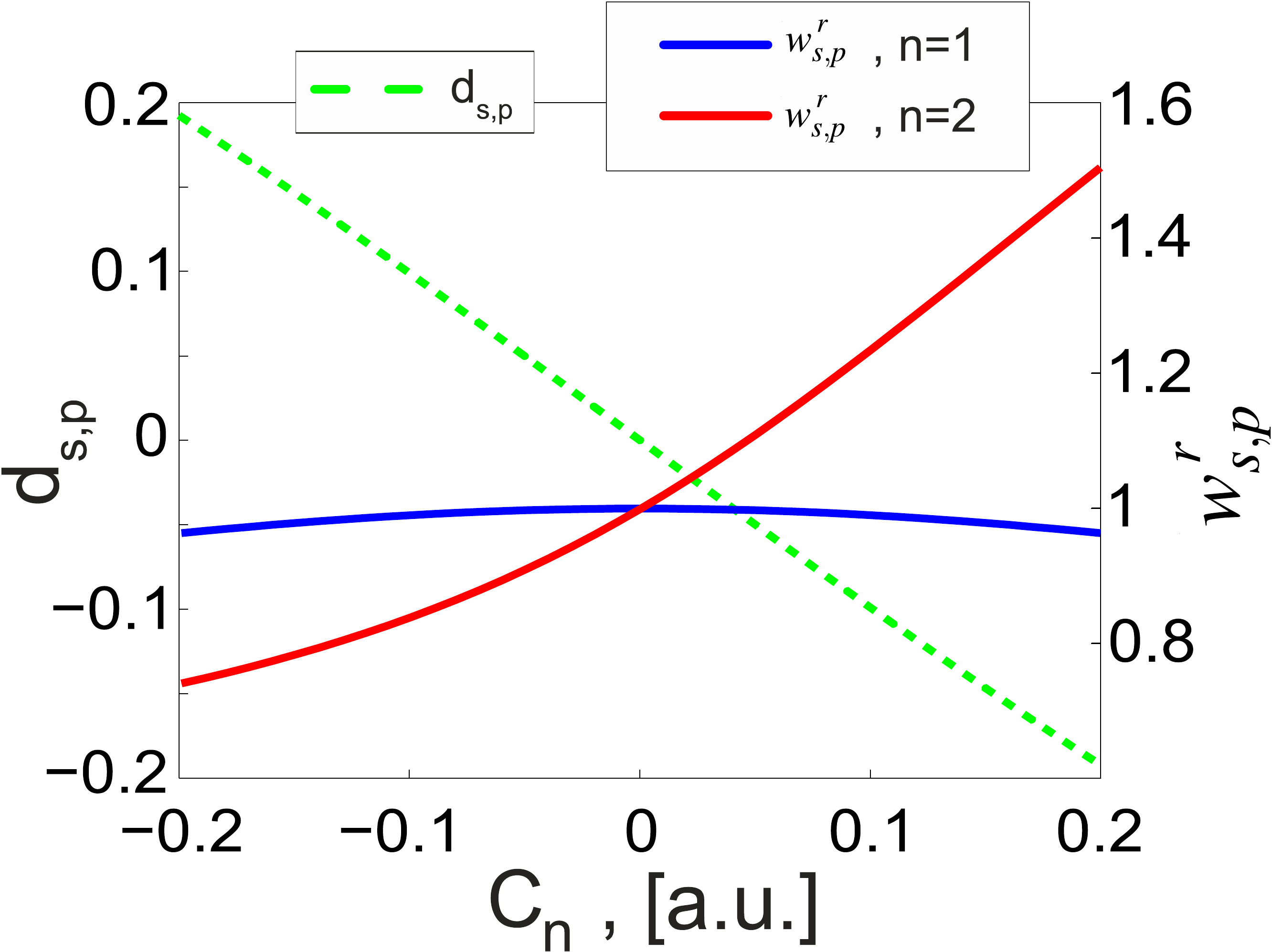} \text{(b)} 
\caption{Schematic depiction of the reconstructed beam profile (black), the electric energy density (red), its first (blue) and second derivative (green) for knife-pads interacting with the incident field via the local electric field (1,2) or via local gradients (3,4) (a).  Dependence of the absolute beam shift from one edge ($x=0$) $d_{s,p}$ (green dotted curve, $n=1$) and the reconstructed beamwidth $w^{r}_{s,p}$ (blue curve for $n=1$ and red curve for $n=2$) on the coefficients $C_n$, for $w_{s,p}=1$ as focal beam diameter.}
\label{fig:theo1}
\end{figure}

The physical meaning behind Eq. (\ref{eq:adopt_der}) is the following. The first term in the sum ($n=1$) is due to the local response of the knife-pad to the $s$- or $p$-polarized electric field density. The second term ($n=2$) expresses the local response of the knife-pad to the gradient of the electric field density and so on. 

Let us discuss this result by looking at the example of $s$- or $p$- polarized Gaussian beams. We can assume, that the projection of its electric energy density distribution $U_{E}( x_0)$ can be written as $U_{E}( x_0)=\exp \left(-x_0^2/w_{s,p}^2 \right)$, where $w_{s,p}$ is related to the FWHM $W_{s,p}$ by $W_{s,p} = 2\sqrt{\mathrm{ln}2}w_{s,p}$. In this case, the Eq. (\ref{eq:adopt_der}) can be written as
\begin{align}
	\frac{\partial P}{P_0\partial x_0}=  \; &C_0  \exp \left(-x_0^2/w_{s,p}^2 \right) - \frac{2x_0 C_1}{w_{s,p}^2}\exp \left(-x_0^2/w_{s,p}^2 \right) \nonumber \\
	-&\frac{2\left(w_{s,p}^2 - 2 x_0^2\right)C_2}{w_{s,p}^4} \exp \left(-x_0^2/w_{s,p}^2 \right)-...
	\label{eq:adopt_der2}
\end{align}
The first order derivative (second term in equation (\ref{eq:adopt_der2})) containing a polarization-dependent amplitude introduces an apparent shift of the effective reconstructed beam projection, see Fig. \ref{fig:theo1} (a), this result is a trivial manifestation of $U_E(x+dx)=U_E(x)+U'_E(x)dx$. Therefore the knife-pad can interact with the local intensity in such a way that the measured beam profile is shifted away from the knife-pad (see Fig. \ref{fig:theo1} (a),  inset 1), while for other interaction scenarios the beam profile can be shifted towards the knife-pad (see, inset 2). The appearance of the second order derivative (third term) in the equation (\ref{eq:adopt_der2}) introduces changes to the beamwidth of the reconstructed beam projection, see Fig. \ref{fig:theo1} (a), and can be attributed to a dilation (scale) operator acting on the beam profile $U_E(x[1+\delta])$, with $\delta$ being a small scaling factor. Here the knife-pad can interact with the local gradients of the field in such a way that the reconstruction results in a broader beam profile (see Fig. \ref{fig:theo1} (a), inset 3), whereas under other conditions the measured apparent beam profile can also be smaller than the real value (inset 4). We note, that the strong response to the local electric field density is causing not only a shift of the beam profile, see Fig. \ref{fig:theo1} (b) (green dotted line), but also slightly changes the beamwidth (blue line). The second order effect is the main cause for the changes in the beamwidth of the reconstructed beam, see Fig. \ref{fig:theo1} (b) (red line). If the coefficient of the second order derivative has a large positive value, the reconstructed beam is not only larger than the real value, but also a strong distortion may appear, drastically altering the shape of the beam profile. At the same time, the measured normalized beam projections might also take values below zero for larger values of $C_1$ and $C_2$. 

Next, we will investigate the deeper meaning behind the expansion coefficients $C_n$ accounting for different physical effects. We will separately investigate effects introduced by the knife-pad and it's edges.

\subsection{Effects of the metal pad}
We will consider here only the conventional term of the knife-edge method leaving effects happening on the edge for the next subsection. We also assume here, that some part of the energy could penetrate the metal pad from the top, therefore we formally extend the integration region in Eq. (\ref{eq:knife}) and rewrite it as
\begin{align}
P = P_0\int _{-\infty}^{\infty} \mathrm{d}x \int _{-\infty}^{\infty}\mathrm{d}k_x\hat{U}_E(k_x,x_0)\hat{T}(k_x)\mathrm{e}^{\mathrm{i}k_xx} .
	\label{eq:knife_ext2}
	\end{align}
Here, $\hat{U}_E(k_x,x_0)$ is the Fourier-image of the signal $U_E(x_0)$, which we expect to measure and $\hat{T}(k_x)$ is a spectral representation of the polarization-dependent knife-edge interaction operator. Due to the integration performed over the $y$-axis we can consider independent classes of $2D$-solutions: transverse electric (in our notation $p$-polarized) and transverse magnetic ($s$-polarized) modes. The incident field in the spectral domain is represented as plane waves with amplitudes $S\left(k_x \right)$ traveling at different angles $\alpha = \arcsin k_x/k$. Here $k=\omega/c_0$ is the wave-vector, with $k_x$ and $k_z$ being the transverse and longitudinal components of the wave vector, $\omega$ is the frequency and $c_0$ is the speed of light in vacuum. 

\begin{figure}[t!]
\centering
\includegraphics[scale=0.20]{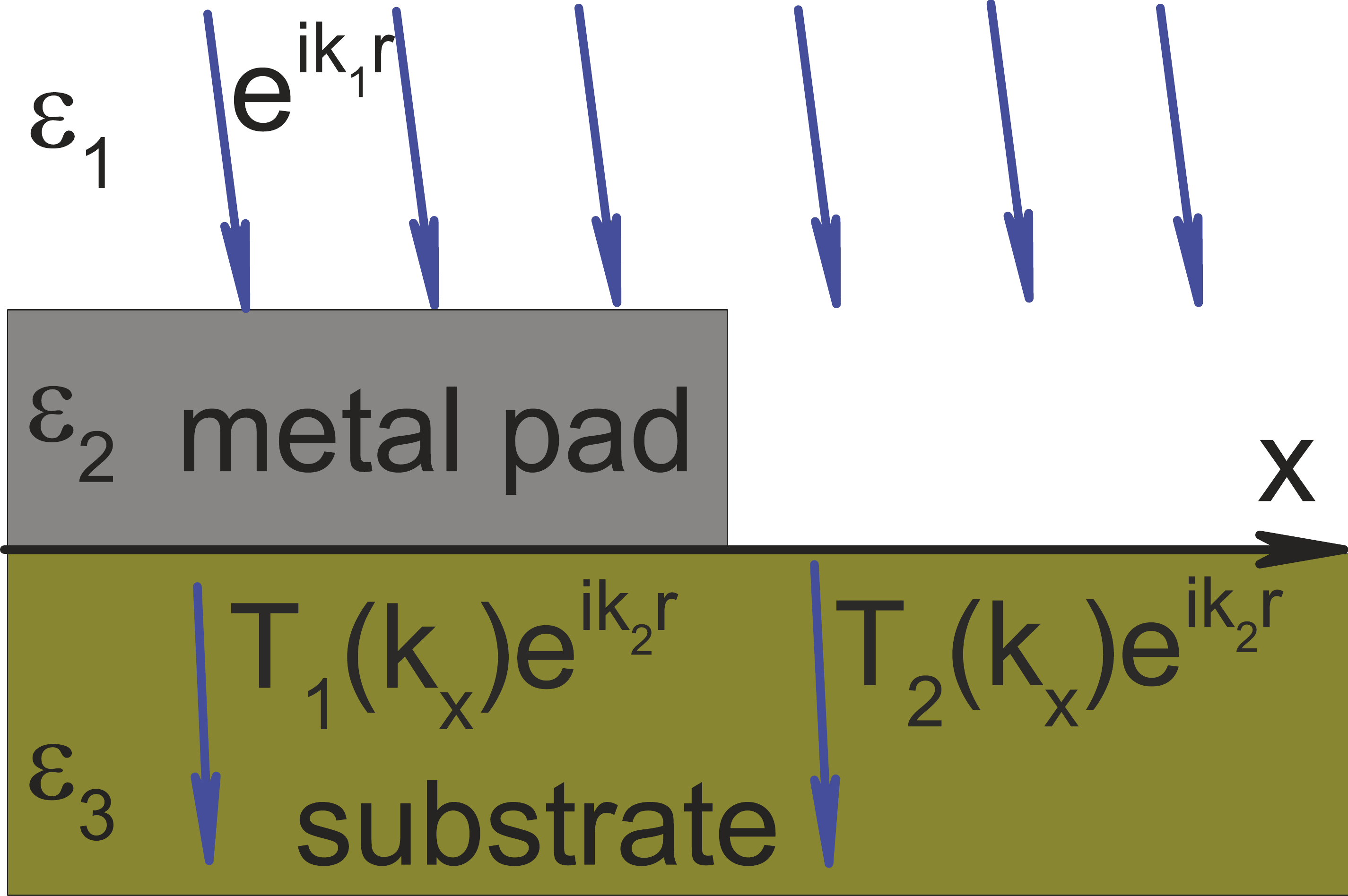} \text{(a)} 
\includegraphics[scale=0.21]{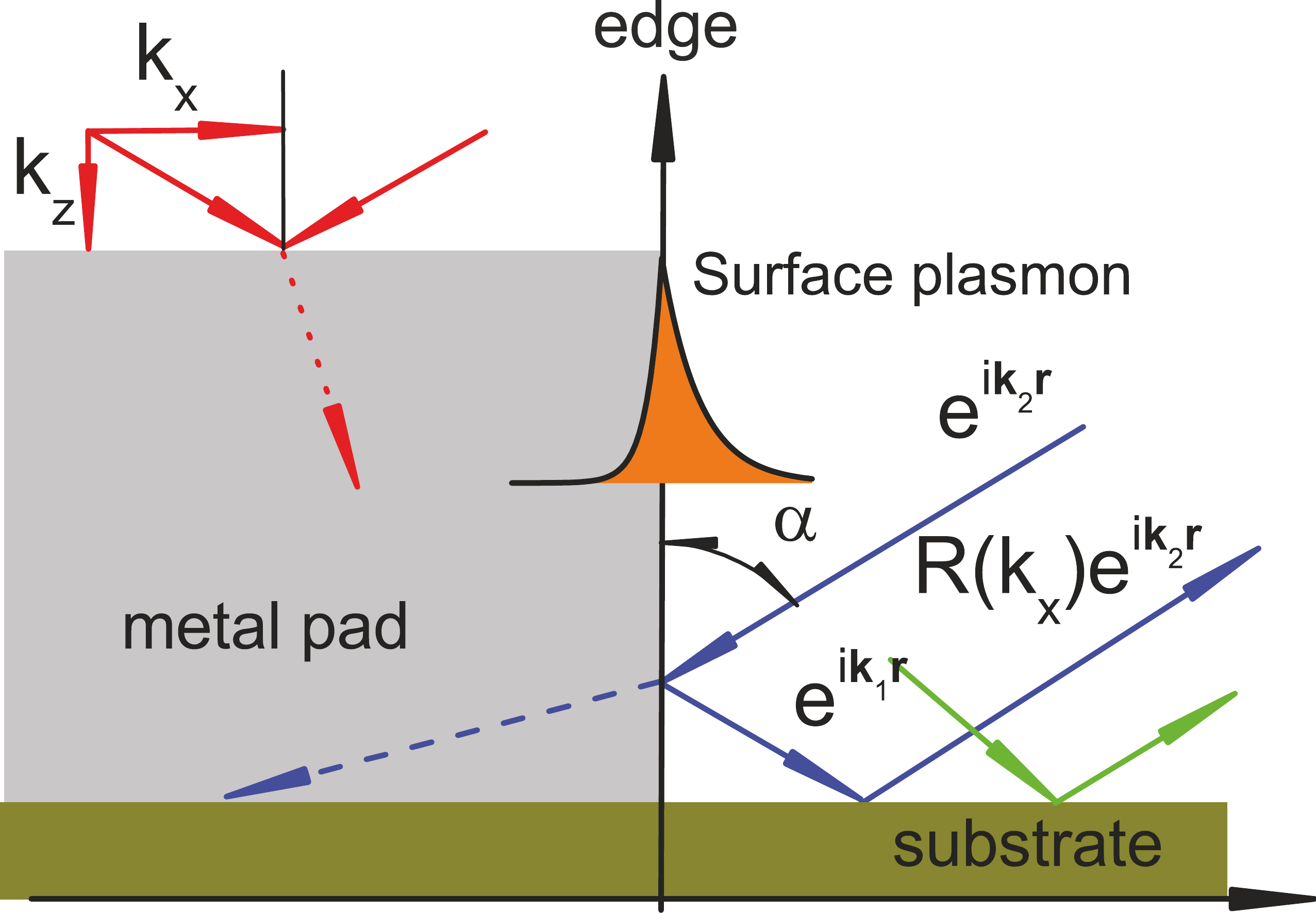} \text{(b)} 

\caption{Sketch of the considered structure containing the medium with dielectric constant $\epsilon _1$, the knife-pad (dielectric constant $\epsilon _2$) and the substrate (dielectric constant $\epsilon _3$). Schematic illustration of a single plane wave component $\mathbf{k}=\left(\pm k_x,k_z\right)$ impinging on the metal pad and being transmitted into the substrate through it or directly (a). Schematic depiction of a single plane wave component ($\mathbf{k}_1=\left(k_x,k_z \right)$, $\mathbf{k}_2=\left(-k_x,k_z \right)$) impinging on the edge and $\mathrm{r}=\left(x,z\right)$. A plasmonic mode is also schematically depicted (b). }
\label{fig:theo2}
\end{figure}

We start our investigation by considering a single plane wave component $\mathbf{k}_1=\left(k_x,k_z \right)=k\left(\pm  \sin \alpha, -\cos \alpha \right)$ of the spatial spectrum $\hat{U}_E(k_x,x_0)$. We consider here a simplified situation, where the part of the plane wave impinging on the metal pad (i.e. for $x \in (-\infty,0]$) enters the substrate with a polarization-dependent intensity $T_1(k_x)$ and the part of the plane wave directly impinging on the substrate (i.e. for $x \in [0,\infty)$) enters with an intensity $T_2(k_x)$, see Fig. \ref{fig:theo2} (a). The function $T_1(k_x)$ is the standard Fresnel transmittance coefficient through stratified media of the height $h$ and the function $T_2(k_x)$ is the corresponding formula for transmittance into the substrate. We assume that both parts are properly detected by a detector. Therefore, we can rewrite Eq. (\ref{eq:knife}) approximately as
\begin{align}
P = P_0\int _{-\infty}^{0} \mathrm{d}x \int _{-\infty}^{\infty}\mathrm{d}k_x\hat{U}_E(k_x,x_0)T_1(k_x)\mathrm{e}^{\mathrm{i}k_xx}+ 
P_0\int _{0}^{\infty} \mathrm{d}x \int _{-\infty}^{\infty}\mathrm{d}k_x\hat{U}_E(k_x,x_0)T_2(k_x)\mathrm{e}^{\mathrm{i}k_xx}.
	\label{eq:knife3}
	\end{align}
We note, that both terms in Eq. (\ref{eq:knife3}) look similar to the standard expressions of the knife-edge method except for two additional functions which modify the resulting spatial spectrum. Indeed, the first term behaves like a knife-pad blocking the region of $x \in (-\infty,0]$ and the second term behaves like a knife-pad placed at $x \in [0,\infty)$. As a further step let us introduce the Taylor expansion of $T_{1,2}(k_x)$  with 
\begin{equation}
	T_{1,2}(k_x) = T_{10,20} + \sum_{n=1}^{\infty}\frac{k_x^n}{n!}\left.\frac{\partial^n T_{1,2}(k_x)}{\partial k^n_x}\right|_{k_x=0},  T_{10} = T_{1}(k_x = 0),   T_{20} = T_{2}(k_x = 0).
	\label{eq:trans}
\end{equation}
We note that $\hat{U}_E(k_x,x_0)=\hat{U}_{E}(k_x)\mathrm{e}^{ \mathrm{i}k_xx_0} $, where $\hat{U}_{E}(k_x)$ is the Fourier image of the electric energy density projection $U_{E}(x)$ of the beam exactly at the position of the knife-pad ($x=0$). We substitute Eq. (\ref{eq:trans}) into Eq. (\ref{eq:knife3}). We use the relation $\partial ^n U_E(x)/\partial x^n = \int \mathrm{d}k_x \left(\mathrm{i}k_x\right)^n\hat{U}_E\left(k_x \right)\mathrm{e}^{\mathrm{i}k_xx}$, so the resulting expression reads
\begin{align}
	P=P_0\int _{-\infty}^{x_0}\mathrm{d}x \left[ T_{10}U_E(x)+ \sum_{n=1}^{\infty}  A_n\frac{\partial ^n U_E(x)}{\partial x^n}\right] +P_0\int _{x_0}^{\infty}\mathrm{d}x \left[T_{20} U_E(x)+ \sum_{n=1}^{\infty}  B_n\frac{\partial ^n U_E(x)}{\partial x^n}\right] ,
	\label{eq:adopted}
\end{align}
with $A_n = (\mathrm{i}^nn!)^{-1}\partial^n T_1/\partial k^n_x$ and $B_n = (\mathrm{i}^nn!)^{-1}\partial^n T_2/\partial k^n_x$. Taking the derivative of Eq. (\ref{eq:adopted}) results in
\begin{align}
	\frac{\partial P}{P_0\partial x_0}=\left(T_{10}-T_{20} \right)U_{E}(x_0) + \sum_{n=1}^{\infty}\left(A_n-B_n \right)\frac{\partial^n U_{E}( x_0)}{\partial x_0^n}.
	\label{eq:adopt_der_tr}
\end{align}

Therefore, physical effects introduced by a substrate and a stratified metal of height $h$ contribute to the identification of the expansion coefficients $C_n$ in Eq. (\ref{eq:adopt_der}) as
\begin{equation}
 C_0 = T_{10} - T_{20}, \quad C_{n} = \delta _{n,2m}(\left. A_n- B_n\right)
	\label{eq:C_n_one}
\end{equation}
where $\delta _{n,2m}$ is Kronecker's delta, $m$ is an integer number and for odd $n$ we have $C_n=0$. Eq. (\ref{eq:C_n_one}) shows, that for the effects taken into account so far the metal pad does not introduce shifts into the beam profile, it accounts only for symmetric distortions. Next, if the height $h$ of the metal pad is larger than the depth of the skin effect, the metal pad itself does not contribute to the transmission, i.e. $ T_{10}=0$ and $A_n =0$. In this case the appearance of distortions in the conventional knife-edge method is governed by the modifications to the spatial spectrum which are introduced by the substrate alone. Lastly, if the substrate introduces no angular dependence or the Fresnel coefficient $T_2 (k_x)$ of the substrate can be neglected (i.e. $B_n=0$ for $n>1$), than Eq. (\ref{eq:adopt_der_tr}) is transformed into the expression from the paraxial knife-edge theory, compare with Ref. \cite{AHFir77}.
\subsection{Effects of the edge}
In this subsection we will study nonconventional effects occurring directly at the edge of the structure, i.e. effects observed at the vertical surface of the knife-pad. In Fig. \ref{fig:theo2} (b), three possible interaction scenarios for a plane wave hitting the edge of the knife are shown in different colors. First, the part of the plane wave spectrum, represented by the red rays, which is either transmitted into the knife-pad or reflected from it. Second, the partial ray shown in green, which is affected only by the substrate. These parts were considered in the previous subsection. Furthermore, the part shown in blue, which penetrates the vertical surface of the knife-pad respectively experiences reflection from it and can enter the substrate afterwards. Lastly, for a spectrum of plane waves forming the spatially confined focal field under study there is a position dependent plasmonic excitation of the knife-pad, which can account for up to 90\% of the total transmission, see \cite{PMar11}.

We consider in what follows those plane waves depicted in blue, which penetrate the vertical surface of the knife-pad. We assume a finite conductivity of the verctical surface and restrict ourselves to a nearly ideal reflection $R \approx 1$ (see Fig. \ref{fig:theo2} (b)). The boundary conditions for $s$- and $p$-polarizations at the side-wall can be written as $\epsilon_1 E_x^{(1)} = \epsilon_2 E_x^{(2)}$ and $E_y^{(1)} = E_y^{(2)}$, where the electric field components are denoted by superscripts. Due to the fact that the fields in the knife-pad decay exponentially as $\exp \left(-\kappa kx \right)$, where $\kappa= \mathrm{Im} \epsilon_2 ^{1/2}$, they contribute to the total detected power $P$ as a term $T_{10}U_E(x_0)\frac{\mathrm{Re}\upsilon}{\kappa k}$, where $\upsilon = \frac{\epsilon_1}{\epsilon_2}$ for a TM incoming field and $\upsilon = 1$ for the TE field. The differentiation of the photocurrent signal in transmission gives
\begin{equation}
\frac{\partial P_{trans}}{\partial x_0}= T_{10}\frac{\partial U_E(x_0)}{\partial x_0}\frac{\mathrm{Re}\upsilon}{\kappa k},
\label{eq:dTrans1}
\end{equation}
and can be interpreted as the response of the edge to the local field intensity resulting in positive or negative shifts for $s$- and $p$-polarized light, respectively as discussed above. The parameter $\upsilon$ depends on the polarization relative to the edge and the substrate material enters via $T_{10}$. From Eq. (\ref{eq:dTrans1}) we can identify the expansion coefficients $C_n$ as $C_0= 0$, $C_1=T_{10}\mathrm{Re}\upsilon/(\kappa k)$ and $C_n=0$ ($n>1$).

Next, we consider the part which is reflected from the vertical surface and transmitted into the substrate. We assume now an angle-dependent reflection $R\left(k_x \right)$ (see Fig. \ref{fig:theo2} (a)). For the two cases $s$- and $p$-polarization, the part of the signal reflected from the side-wall and entering the substrate can be expressed as
\begin{align}
P_{ref} & =  \int _{0}^{\infty} \mathrm{d}k_x \int _{0}^{\frac{hk_x}{\sqrt{k^2-k_x^2}}}\mathrm{d}x\hat{U}_E(-k_x,x_0)\frac{R\left(\sqrt{k^2-k_x^2}\right)}{\sqrt{k^2-k_x^2}}T_2(k_x)\mathrm{e}^{\mathrm{i}k_xx} \nonumber \\
 & = \int _{0}^{\infty} \mathrm{d}k_x \hat{U}_E(-k_x,x_0)\frac{R\left(\sqrt{k^2-k_x^2}\right)}{\sqrt{k^2-k_x^2}}T_2(k_x)\frac{\mathrm{e}^{\frac{\mathrm{i}hk_x^2}{\sqrt{k^2-k_x^2}}}-1}{ik_x},
	\label{eq:reflect}
\end{align}
where $R(k_x)$ is the standard reflectivity coefficient and $h$ is the height of the knife-pad. After differentiation of the Eq. (\ref{eq:reflect}) we can approximate the unknown $C_n$ coefficients as
\begin{equation}
 C_0 = 0, \quad  C_n = -\frac{1}{\mathrm{i}^nn!}\left. \frac{\partial^n }{\partial k^n_x}\left\{\frac{T_2\left(k_x \right)R\left(\sqrt{k^2-k_x^2}\right) k }{i k_x \sqrt{k^2-k_x^2} }\left[\exp \left({\frac{\mathrm{i}hk_x^2}{\sqrt{k^2-k_x^2}}}\right)-1 \right]\right \}\right|_{k_x=0} .
	\label{eq:C_n_two}
\end{equation}
We see, that the reflection from the edge of the knife-pad only introduces effects described by first and higher orders of the expansion coefficients $C_n$. Here we have neglected the part of the signal entering the knife-pad from the top and leaving the knife-pad from the vertical surface, because plane waves exit the system as inhomogeneous evanescent plane waves (see Fig. \ref{fig:theo2} (a), red color).
In a last step, we investigate the influence of the plasmonic excitation \cite{Maier} of the knife-pad. For the sake of simplicity we assume that the incident field only excites plasmonic modes of the knife-pad. This way, we avoid rather complicated boundary conditions and an interplay between plasmonic and non-plasmonic modes of the knife-pad \cite{HLOff05, YZha13}. It is worth noting here, that this step only grants access to the physical effects underlying the interaction process, but does not allow for a quantitative study. For an overview of the complete theoretical model we redirect the reader to Ref. \cite{PMar11}.

The strength with which a plasmonic mode is exited can be estimated from an overlap integral of the projection of the incident electric field and the magnetic field of the plasmon (Fig.  \ref{fig:theo2} (b))
\begin{equation}
L= \epsilon^{*}_2\int _{-\infty}^{0} \mathrm{d}x E_b\left(x+ x_0 \right)\exp \left( -\mathrm{i}\kappa_2^{*} x \right) + \epsilon^{*}_1\int _{0}^{\infty}\mathrm{d}x E_b\left(x+ x_0 \right) \exp \left( -\mathrm{i}\kappa_1^{*} x \right),
\label{eq:L1}
\end{equation}
where \cite{Maier}
\begin{equation}
\beta= k\sqrt{\frac{\epsilon_1 \epsilon_2}{\epsilon _2 +\epsilon _1}}, \quad \kappa_1 = \sqrt{k^2 \epsilon_1 -\beta ^2}, \quad \kappa_2 = \sqrt{k^2 \epsilon_2 -\beta ^2}.
\label{eq:bet}
\end{equation}
We rewrite Eq. (\ref{eq:L1}) in Fourier domain as
\begin{equation}
L= \epsilon^{*}_2 \int _{-\infty}^{\infty}  \mathrm{d}k_x \frac{ -\mathrm{i}\hat{E}_b\left(k_x \right)}{-k_x + \kappa_2^{*}} \exp \left( \mathrm{i} k_x x_0 \right) + \epsilon^{*}_1\int _{-\infty}^{\infty}  \mathrm{d}k_x \frac{-\mathrm{i} \hat{E}_b\left(k_x \right)}{k_x - \kappa_1^{*}} \exp \left( \mathrm{i} k_x x_0 \right),
\label{eq:L2}
\end{equation}
and after performing a Taylor expansion we arrive at the following expression
\begin{equation}
L= \sum _{n=1}^{\infty} c_n\frac{\partial ^{n-1} E_b(x_0)}{\partial x_0^{n-1}}, \quad c_n= \frac{1}{\mathrm{i}^{n-1}} \left[\frac{\epsilon^{*}_2 }{(\kappa_2^{*})^{n+1}} - \frac{\epsilon^{*}_1 }{(\kappa_1^{*})^{n+1}} \right] .
\label{eq:L3}
\end{equation}
A distinct feature of Eq. (\ref{eq:L3}) is that it contains no term with $n=0$ due to $\epsilon_1/\kappa_1 = \epsilon_2/\kappa_2$, meaning that there is no term corresponding to a simple blocking of the beam.
The power transmitted into the substrate is proportional to the square of the plasmon amplitude $|L|^2$ and differentiating  $|L|^2$ with respect to $x_0$ results in $d|L|^2/dx_0=L^{*}dL/dx_0 +LdL^{*}/dx_0$. Keeping in mind that $U_{E}(x_0)= E_b\left( x_0 \right)E_b^{*}\left( x_0 \right)$ and that we can use Leibniz rule to find $d^nU_{E}(x_0)/dx_0^n$, it becomes obvious, that the expression for $|L|^2$ can be recast into a sum of $d^nU_{E}(x_0)/dx^n_0$, similar to Eq. (\ref{eq:adopt_der}) and the coefficients $C_n$ can be identified as
\begin{equation}
C_n= \sum _{m=0}^n c_mc^{*}_{n-m}\frac{n!}{m!(n-m)!},  \text{ for }  n>1 .
\label{eq:C_n_three}
\end{equation}

\subsection{Application of the simplified model to the experimental situation}
We start by demonstrating some measured beam profiles to visualize asymmetric distortions of these curves. For BK7 as substrate material and p-polarization two differentiated photocurrent curves are compared for two wavelengths of $700$ nm and $535$ nm.
While the left side of the profile ($x<0$) quite nicely resembles a Gaussian shape, the right side of the profiles decay more slowly and behave more like an exponential function. This can be seen well especially for smaller wavelengths (see curve for $\lambda = 535$ nm), which is close to the plasmonic resonances of gold. 

\begin{figure}[t!]
\centering\includegraphics[scale=0.75]{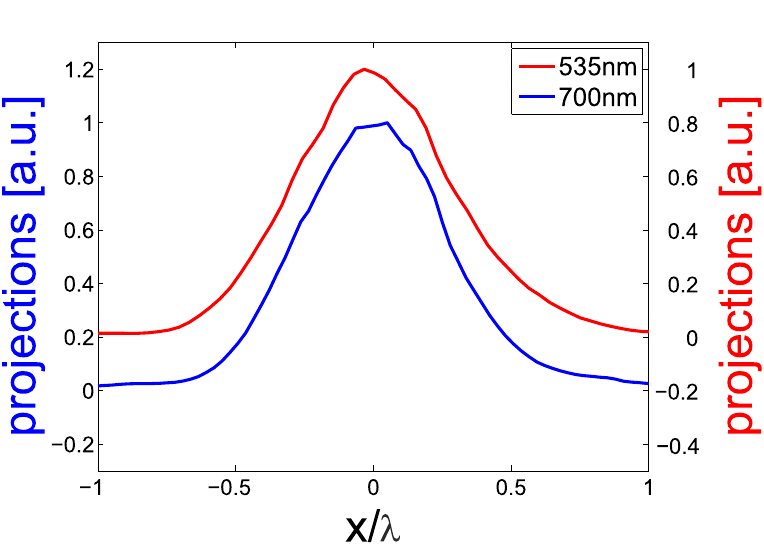}\text{(a)}
\centering\includegraphics[scale=0.2]{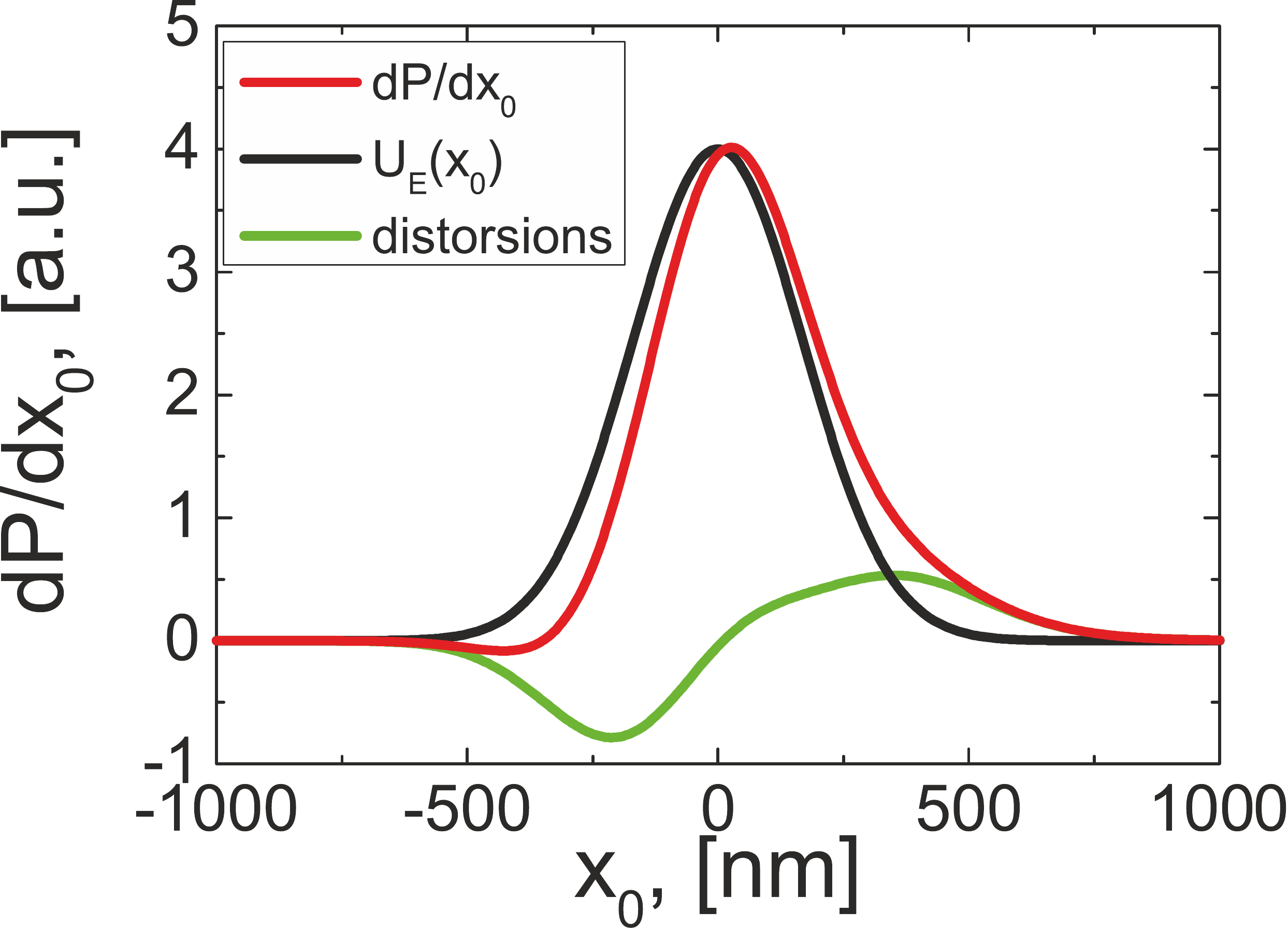}\text{(b)}
\caption{Experimentally measured beam profiles for $p$-polarization and two wavelengths ( $\lambda = 700$  nm, $535$  nm). The measurements have been performed using a knife-pad with a height of $70$ nm placed on a BK7 glass substrate (a). Numerically calculated derivative of the photocurrent $\partial P /\partial x_0$ (red), expected beam profile $U_E(x_0)$ (black) and a term representing distortions ($\partial P /\partial x_0 - U_E(x_0)$)  to the beam profile (green) for one particular wavelength $\lambda = 535$  nm (b). }
\label{fig:profilesplasmon}
\end{figure}

An appearance of the exponential part might appear surprising, but it can be numerically confirmed using our exact theoretical model \cite{PMar11}, see Fig. \ref{fig:profilesplasmon} (b). We have numerically calculated the derivative (red color) for one particular wavelength ($\lambda = 535$  nm) for the same sample as in Fig. \ref{fig:profilesplasmon} (a) and one can easy note a strongly asymmetric shape of the beam profile. Due to the presence of an exponential tail, the corrected knife-edge technique introduced in Ref. \cite{CHub13} requires the involvement of a large number of high order derivatives.

We can explain an appearance of the exponential tail using simplifications from the last subsection. We assume again that the electric field of the beam can be expressed as $E_b( x)=\exp \left(-x^2/w_{s,p}^2 \right)$. In this case, the Eq. (\ref{eq:L1}) can be expressed as
\begin{eqnarray}
L= &\epsilon^{*}_1w_{s,p}\sqrt{\pi}/2 \exp \left[-\frac{\kappa_1^{*}}{4}\left(\kappa_1^{*}w^2_{s,p}-4\mathrm{i}x_0 \right) \right] \left[1+\mathrm{erf}\left(i\kappa_1^{*}w_{s,p}/2+x_0/w_{s,p} \right) \right] \nonumber \\
+ &\epsilon^{*}_2w_{s,p}\sqrt{\pi}/2 \exp \left[-\frac{\kappa_2^{*}}{4}\left(\kappa_2^{*}w^2_{s,p}-4\mathrm{i}x_0 \right) \right] \left[1-\mathrm{erf}\left(i\kappa_2^{*}w_{s,p}/2+x_0/w_{s,p} \right) \right].
\label{eq:L1Gauss}
\end{eqnarray}
For rather large displacements $x_0>0$, Eq. (\ref{eq:L1Gauss}) can be further simplified because the erf-function takes values close to one and thus we can write
\begin{eqnarray}
L \approx &\epsilon^{*}_1w_{s,p}\sqrt{\pi} \mathrm{e}^{-\left(\frac{\kappa_1^{*}w_{s,p}}{2}\right)^2} \mathrm{e}^{\mathrm{i}\kappa_1 x_0} .
\label{eq:L2Gauss}
\end{eqnarray}
This means, that if the beam is being moved rather far away from the knife-pad, the plasmon excitation is proportional to an exponential function of displacement $x_0$, see Eq. (\ref{eq:L2Gauss}). Taking a derivative of $|L|^2$ will result in a term proportional to $\exp \left(2 \mathrm{Im} \kappa_1 x_0 \right)$, therefore an appearance of an exponential tail in Fig. \ref{fig:profilesplasmon} can be explained by this plasmonic excitation. Although the beam position $x_0$ influences the plasmonic excitation via the amplitude $L$ the substrate material enters into equations, see Ref. \cite{PMar11}, through the boundary conditions. For GaAs and Si substrates these boundary conditions result in a smaller effective influence of this plasmonic excitation on the measured beam profile.
Next, our numerical results and the simplified model from the last section predict that an $s$-polarized beam is shifted away from the knife-pad whereas an $p$-polarized beam is shifted into the knife-pad due to the different interaction of the beam with the knife-pad for most cases. In principle, it is sophisticated to determine the widths of the knives used in the experiment by SEM measurements with sufficient accuracy to be able to determine this absolute shift by experiment. At last, our numerical calculations and our simplified considerations also predict, that the beam sizes of the measured projections should be modified due to the interaction of the knife-pad with the electric field and in special with its local gradients regarding $C_2$. Experimental results confirm those predictions in all considered cases.

\section{Conclusion}
In conclusion, we have extended our previous investigations on the knife-edge method \cite{PMar11} by considering the influence of the substrate of the knife-edge samples, which was left unaccounted for so far. The choice of the substrate has a crucial impact on the measured projections of the beam. Experimental results for three different substrate materials and two heights of the knife-pads are discussed and compared to our exact calculations. Moreover, based on our previously introduced adapted analysis approach of knife-edge data \cite{CHub13}, we present a simple physical scheme, which explains how underlying physical effects are influencing knife-edge measurements and what causes both beam shifts and distorted beam profiles.

\section{Acknowledgements }
We thank Stefan Malzer, Isabel G\"assner, Olga Rusina and Irina Harder for their valuable support in preparing the samples. Peter Banzer acknowledges the support by the Alexander von Humboldt Foundation.
\end{document}